# WEATHER SEQUENCES FOR PREDICTING HVAC SYSTEM BEHAVIOUR IN RESIDENTIAL UNITS LOCATED IN TROPICAL CLIMATES.

L. Adelard, F. Garde, F. Pignolet-Tardan, H. Boyer, J.C. Gatina

Laboratoire de Génie Industriel, Université de la Réunion, 15, Avenue René Cassin.

97715 Saint Denis Cedex 9 Reunion Island, France. Tel. 262 93 82 23

Email: adelard@univ-reunion.fr

## ABSTRACT

The purpose of our research deals with the description of a methodology for the definition of specific weather sequences and their influence on the energy needs of HVAC system. We'll apply the method on the tropical Reunion Island. The methodological approach based on a detailed analysis of weather sequences leads to a classification of climatic situations that can be applied to the site. These sequences have been used to simulate buildings and air handling systems thanks to a thermal simulation code, CODYRUN. Results bring to the light how necessary it is to have coherent meteorological data for this kind of simulation.

## INTRODUCTION

Reunion island is situated by +21° South Latitude and 55° East longitude, next to Madagascar island. Situated in a tropical zone, the year is theoretically divided in two seasons:

- In the humid season (from November to April), the island is close to the inter tropical convergence zone and cyclones perturbations may occur.

The fresh season is from May to October, when the climate of the island is influenced by trade winds. The island is highly mountainous. There are smooth bents on the cost zones, which increase quickly toward the centre of the island. The centre is made of three cirques which give a very contrasted relief. Such complexity on a small surfaces (2500 Km.²), gives a lot of micro climates.

Demography, associated with the importation of unadapted construction models leads to the development of buildings totally unadapted to humid climates. The sensations of thermal discomfort in these buildings have increased the energy consumption by the use of individual air conditioning systems. It was then necessary to develop bioclimatic architecture. This development needs precise information about the effect of climate patterns on buildings.

For energy applications in buildings, meteorological data are needed for simulations. In buildings simulations, the most important climatic situations are due to solar radiation and air temperature. In the Reunion thermal environment, the buildings depend on external climatic conditions. Heat transfers are due to convection, conduction and above all to solar radiation. That is different from the temperate climate zones. Wind speed and air humidity are also important in studies of the human comfort in buildings [1]. It's not the effect of one climatic variable that is important for the building behaviour, but somewhere the combination of the different climatic variables [2].

Meteorological data can be unavailable for different reasons:

- There is no meteorological station for the site.
- Data are missing in long periods.

Database that can represent the main current or extreme climatic situations of the site, for the different periods of the year doesn't already exist. So, the purpose of our research is to first analyse the climate of selected sites to determine the most important climatic variables to build finally a weather data generator to produce weather sequences.

## METHODOLOGY

**a) Description of methods already used to provide weather data.**

Generated data must respect some conditions:
- They must respect the main statistic for each variables (distributions laws, auto correlation).
- They must respect the interactions between the climatic variables.
- They must take into account the geographic and physic environment effects on the climate of the site.

We can find three types of database for the energy buildings simulations:

- Simulations can be made by using a series of year of hourly data.

- Numerous applications use a Typical Meteorological Year of data [3]. This year is a single year of hourly data selected to represent the range of weather paterns that would typically be found in a multi-year dataset. Definition of TMY depends on its satisfying a set of statistical tests concerning the multi-year parent dataset. This method has a major inconvenient. Building conception (choice of materials, architecture, detached houses, or lodgings) leads to different sensitivity to climatic variables. So, multiple TMY are needed for different priority of the climate elements.

- The «representative days» can be used in simplified simulation and design tools. These days have to represent the typical climatic conditions. They correspond to average conditions for each climate element. They can also be chosen with the help of a statistical classification [4].

- Data can also be generated by a weather data generator. This kind of code generally use statistical analysis to model the climatic variables. Degelman [5] created a generator which used the monthly mean data as entries of the code, so it can be used in all sites. Van Paassen and Dejong [6] also elaborated a weather data generator, his approach was to separate all climatic variables in two parts. The first part is dependent of the solar radiation, and the second part is generated with statistical means.

Instead of generating one year of data, we want to generate climatic sequences, as asked by the user considering its specific needs. If he wants for example a warm sequence for the humid season, our generator must be able to provide the result and to specify if necessary the frequency of this sequence, and the climatic conditions in which this may occur. For example, we won't have the same humidity before or after two rainy days for the same radiation conditions.
Our purpose is to generate not only traditional climatic sequences, but we also want to be able to simulate non current situations. We will use the principles used by Degelman, and Van Paassen to establish the interactions between the climatic variables.

**b) First stage of our method: Identification of the weather conditions.**

Our aims is first to make an inventory of all the climatic conditions that can occur. We begin then to represent the daily evolution of each variables by specific daily indicators (table 1). These indicators were chosen in function of the utilisation of the users of simulation tool.

Table 1: Definitions of the indicators.

| Variables | Indicators |
|---|---|
| Temperature | Daily maximum Daily mean |
| Humidity | Daily mean Daily minimum |
| Wind speed | Diurnal mean Nocturnal mean |
| Wind direction | Daily mean Nocturnal mean |
| Nebulosity | Daily mean |
| Insulation | Daily sum |
| Global radiation | Daily sum |
| Diffuse radiation | Daily sum |

In humid season, the daily mean temperature does not vary a lot ( from 24 to 28C). That's the reason why, we preferred to observe the evolution of temperature by his daily maximum, witch is more representative of the variability of temperature. We observed that wind speed decrease in the night (from 21 hours). So, its is more realistic to take only the diurnal mean to represent the evolution of the wind.
We analyse the distribution of each indicators for the different periods of the year. The domain of variation of each indicator is separated into few sections. Each section correspond to a fixed criteria. Figure 1 explains the methods, and table 2 shows the criteria chose for the global solar radiation. We present the results for the station of Gillot. We chose this station because its climate is representative of the climate of the Capital of the island. We can now make some observations about the distributions of each section for the different periods of the year (fig.2). In humid season, multiple situations of daily radiation can appear (very low to very high). In the fresh season, there is a lot of breezy days. In the fresh season, you find more days with medium or strong winds, with a direction of East to North- East. That's the influence of trade winds.

We recorded the days by fixing criteria based on more than one climatic indicators. For example, the users can wish to simulate the buildings with days with high radiation and breeze. We create then little database for each of these criteria. The table 3 shows the classes obtained for the criteria based on solar radiation and wind and figure 3 shows the diversity of climatic situations you can find in humid season.

We make an inventory of currently or extreme climatic situations for the periods of the year. For example, in humid season it is common to find high radiation with breeze (0 to 3 ms$^{-1}$), high radiation with medium wind (from 3 to 6 ms-1). An extreme situation can be a high radiation day with very high relative humidity (form 75 to 85 %).

All these sequences have been reported in a catalogue used by the architects to chose climatic conditions in which they are going to simulate buildings.

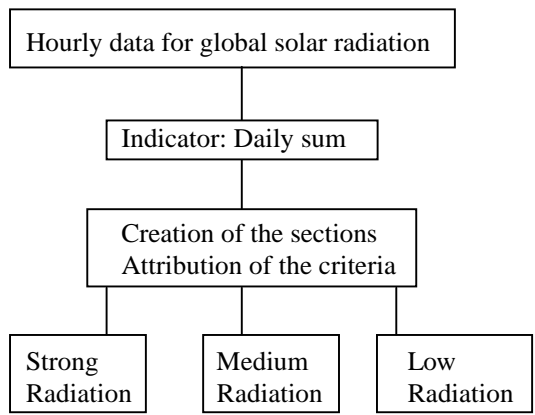

Fig.1: Description of the methodology to determine the criteria

table 2: classes and designations for the global solar radiation.

| Intervals | Designation. |
|---|---|
| from 600 to 2300 Wh.m-2 | very low radiation |
| from 2300 to 4000 Wh.m-2 | low radiation |
| from 4000 to 5700 Wh.m-2 | average radiation |
| from 5700 to 7400 Wh.m-2 | high radiation |
| from 7400 to 8500 Wh.m-2 | very high radiation |

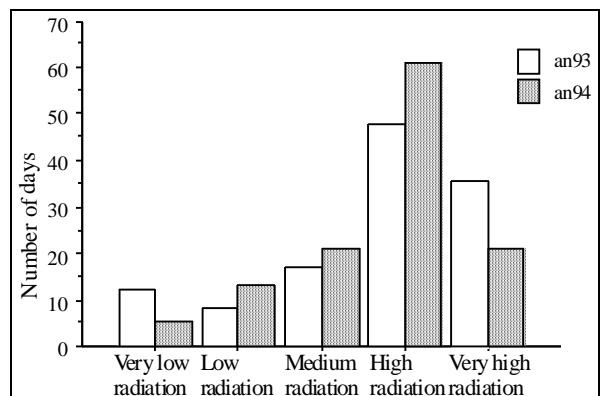

Fig.2: Distribution of global solar radiation for the two humid seasons for the station of Gillot

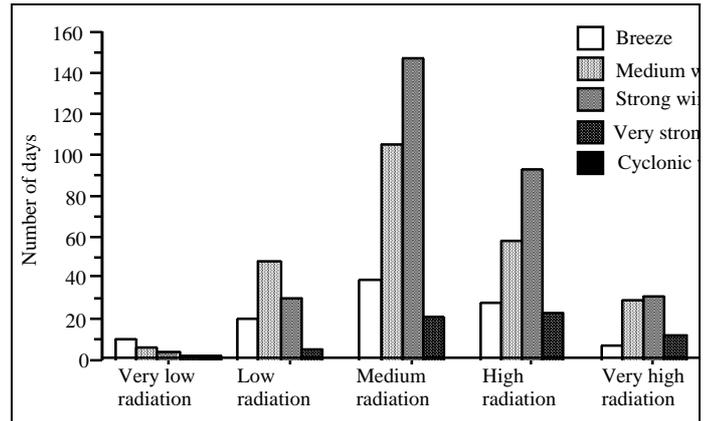

Fig.3: Distribution of wind classes for the different classes of radiation for the years 1993 and 1994.

Table 3: Criteria for radiation and wind

| Designation | Number of days |
|---|---|
| Average radiation, breeze. | 39 days |
| High radiation, breeze | 25 days |
| Low radiation, average wind | 45 days |
| Average radiation, average wind | 92 days |
| High radiation, average wind | 55 days |
| Very high radiation, average wind | 28 days |
| Average radiation, strong wind | 136 days |
| High radiation, strong wind | 88 days |
| Very high radiation, very strong wind | 29 days |
| Average radiation, very strong wind | 26 days |
| High radiation, very strong wind | 23 days |
| Very high radiation, very strong wind | 13 days |

For each of these sequences, we tried to build regressions between the variables. We followed the works made by Mezino [7] and J-Lam [8]. Linear regressions established between the hourly fraction of insulation, the hourly clearness index, the hourly diffuse index gives good results. But no good linear regressions between the hourly data of other variables have been found. However, there is strong auto correlation for the hourly and daily temperature, the hourly wind speed, the hourly and daily humidity.

So, in the next stage of our paper, we are going to expose the methods used to go further in the analysis of these sequences.

c) **Second stage: Use of more elaborated tools.**

As we know that all the variables have high auto correlation, distributions laws for each classes can be determinate with the objective of building stochastic models. The test used to verify that the distributions laws fits with the data is the Chi-square test.

To elaborate stochastic models, it is necessary to know the statistic distribution of the variables. So, we have to evaluate the statistic distributions of variables such as the solar radiation, the wind, the temperature. A gaussian distribution can be used for the air temperature. Authors use generally Weibul distributions for the wind speed. The Weibul distribution is then transformed to have a normal distribution [9] to build an autoregressive model. For the radiation, it's possible to use the distribution described by Saunier for tropical climates [10].

However, we can compare the distribution of the data using the Kolmogorov-Smirnov test. That's what we made for the different classes. This part shows that the wind makes the temperatures lower, due to the convective heat transfer. The wind also reduce the relative humidity. This part shows that the worst climatic situation for the buildings can be days with breeze and very high radiation. In these case, temperature is high and also relative humidity. However, maximum temperatures is measured in windy days.

To analyse the interactions between the variables, we also used the factorial analysis. Factor analysis permit to find the implicit factors that describe the data. These methods can give graphics representations of the studied variables, according to the interactions between them. There are two methods of factor analysis. The first (ACP: principal components analysis) can treat the hourly data by using the correlation matrix between each hour of data. The data can be grouped in different classes. The second methods (ACM: multiple correspondence analysis) can use contingence tables, and use then frequencies of each classes. This second methods can then show non-linear relations. The principles of this analysis is relatively simple: each variable is consider as a point, with a number of components. For example, if you follow the evolution of temperature at 13 hour for a sequence of 130 days, the vector which represent temperature at 13 hours will have 130 components. If you want to evaluate the interactions, you have now to choose and to calculate a statistic distance between all the points. A more detailed description of these two methods can be found in [11]. As the correlation between the hourly data are weak, we'll use ACP to describe the main daily evolution for the different variables. For example, we show here the typical global radiation evolution for the humid season (fig.4 and fig.5 ). The fig. 5 shows a very frequent evolution in humid season in breezy days, the convective effects of global radiation creates clouds in the afternoon, and gives very humid and hot air. ACM shows that global radiation does not directly influence the air temperature and the relative humidity. So it shows that «history» of these two variables the days before is also very important. This point will be very important when we'll choose our sequences for the simulations.

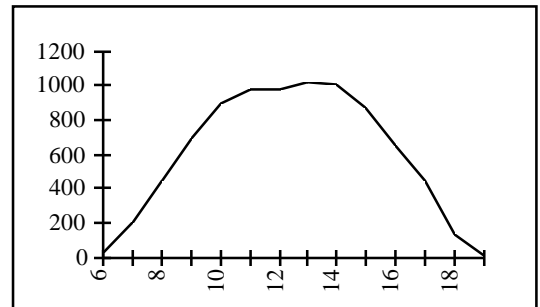

Fig.4: Hourly evolution for the global radiation (in $Wh.m^{-2}$) for a very sunny day.
Daily sum: 8362 $Wh.m^{-2}$

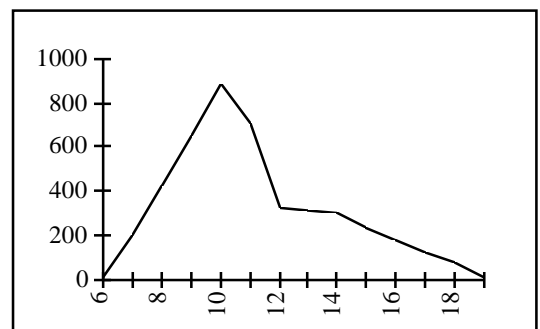

Fig. 5: Hourly evolution for the global radiation (in $Wh.m^{-2}$) for a very sunny morning and cloudy afternoon.
Daily sum: 4408 $Wh.m^{-2}$

That's the reason why we used spectral analysis to study the coherency between variables (Ref. 7). At first, the data of each sequences are filtered to take only the low frequencies to find the daily influences. And then, we calculate the coherency between variables. It shows that solar energy can be considered as an independent variable. The temperature, and the humidity are dependent on solar radiation and also of wind speed when the wind is strong in fresh season.

It would be interesting to evaluate the transfer function between the variables, to evaluate the most influencing variables.

We can then elaborate models that take account of the influence and the history of the variables. For example, the temperature should be generated by using auto correlation and influence of solar radiation.

Another way to establish the relations between the variables is to create some artificial neural networks to generate the data. The network can establish non linear relations and it can take account of all the interactions. It is more and more used in applications of weather predictions [13], but we won't describe the method here.

## APPLICATIONS TO THE BUILDINGS AND HVAC SIMULATIONS.

### a) Description of the software.

The HVAC system being studied will be simulated with an airflow and thermal buildings simulation software, CODYRUN. A description of the model can be found in [14] and [15]. This multizone software has been designed to suit the needs of researchers and professionals. It was developed on a PC micro computer with Microsoft windows user-friendly interface.
CODYRUN integrates both natural ventilation and moisture transfers. It offers to the user a wide range of choices between different heat transfer models. The software is close to ESP [16] for the thermal aspect. The airflow model can be compared to Airnet. The choice of models will depend on the objective of the simulation:
- Study of temperatures regarding to the choice of components of the buildings.
- Estimation of the daily energy consumption.

The software is divided in three parts:
- Module of description:
A systemic building description is made. The building is break down into three parts:
- zones (from a thermal point of view)
- inter-zones ( zones separations)
- components (walls, glass partitions, HVAC systems, ...)
- Module of simulation:
Boyer [14] used the systemic analysis to build mathematical model. So the building is decomposed in few zones. The technique of nodal discretisation of the space variables by finite difference is then used. The sets of equations for the zone thermal model can be condensed in a linear steady state evolution equation:

$$[C] \frac{dT}{dt} = [A] T + B$$

Module of exploitation of results:
The outputs are exported to classical spreadsheets to be exploited.
For the study of HVAC systems, the code includes three types of models. These models have been elaborated from a very simplified one to a very detailed one. A precise description can be found in the paper [17].

The first model is just supposed to supply the demand of cooling loads with an ideal control loop (no delay between the solicitations and the time response of the system). The available outputs are initially the hourly cooling consumption without integrating the real characteristics of the HVAC systems.

The second model is more detailed that the first one. The first step was reduced to one minute in order to take into account the on-off cycling and the control of the systems.
The performances of an air to air heat-pump are strongly influenced by parameters such as the indoor dry air temperature and relative humidity. In the precedent model, they are supposed to be constant.

### b) Presentation of the building.

The studied building is a residential buildings. After consulting the local low cost housing institutions and the new housing statistics for Reunion Island, we found the most common dwelling found is the T3/V, consisting of two bedrooms, a living room and a veranda.

*WEST*

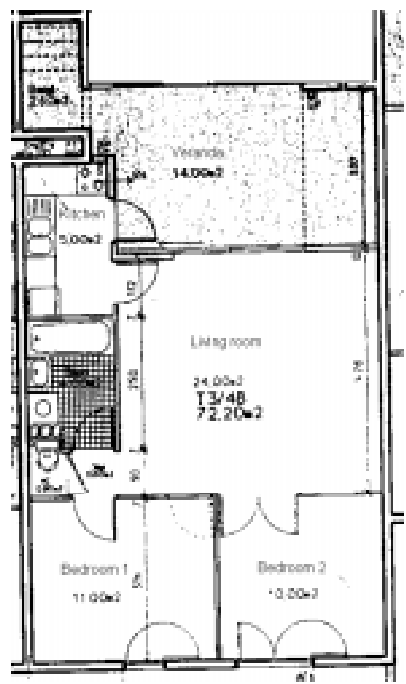

*EAST*
fig.6: Description of the flat used in the simulations.

We are going here to analyse the HVAC energy needs for a flat situated at the top of the building, beneath the roof. The bedroom is oriented towards the East, and the living-room is oriented towards the West. We'll consider the flat as a two zones systems (living-room zone and bedroom zone).

*c)* **Method and Results.**

Haberl and al. [18] showed that it was useful in evaluating energy use to use both measured data and typical meteorological year. We chose several weather sequences of the humid season to simulate HVAC behaviour. We'll follow two ways:
The first way consist in simulating the building by using the three most humid and hot month in the year which are December, January, and February (120 days). We calculate then, as a reference, the mean daily energy consumption for cooling for these months.

Table 4: Evaluation of the consumption with the help of a simulation with the three months of hourly data.

| Sensible cooling consumption (Kwh) | 14 |
| Latent cooling consumption (Kwh) | 3 |
| Total cooling consumption (Kwh) | 17 |

The second way consist in simulating the building by using the weather sequences to analyse precisely the energy consumption in each of these situations.
We have carried out a power study for each of the sequences presented in table 5. Each of these sequences have about 5 days, because the weather varies relatively quickly in the humid season.

For each of these sequence, we give the mean and the maximum energy needed per day. We separate the energy need for dehumidification (the latent cooling capacity) and the energy need for lowering the temperature (the sensible cooling capacity) in the table 6. We represented the mean daily cooling capacities (MEAN), and the maximum daily cooling capacities (MIN).

The results bring to the light the influence of coupling conditions of wind, temperature, humidity, and radiation. The sequence A shows the importance of coupling wind and temperature. This situation requires an important sensible capacity. We thought that latent power would be also important, but it is not the worst situation for the latent cooling needs. We observe then that it is for this case that we find the maximum sensible cooling capacity. The building inertia is showed with this sequence. Effectively, the maximum needs don't appear in the first day, but in the sixth day of the sequence, after five days of breezy and sunny days.

The sequence C contains more hot days with high radiation, but with a medium wind. Even with great convective heat transfer due to the medium wind speed, the sensible cooling capacity is above the mean, however, it is not the most important demand.

For the sequence of high humidity such as sequence E or F, we find the most important demand for the latent cooling capacity. However, we found that for the sequence E, a medium consumption for the sensible cooling capacity which does not exist for the sequence F because The temperature is higher for the sequence E.

Table 5: Description of the sequences.

| Sequences | Description |
|---|---|
| Sequence A | High to very high radiation (from 5700 to 8400 Wh.m-2) Breeze (0 to 3 ms-1) Mean temperature:27°C Maxi temperature:30.7°C |
| Sequence B | High to very high radiation Breeze Mean temperature:25.5°C Maxi temperature:30.8°C |
| Sequence C | High to very high radiation Medium to strong wind (from 3 to 9 m.s$^{-1}$) Mean temperature: 28°C Maxi temperature:33.2°C This sequence is current when a cyclonic perturbation is in proximity of the island |
| Sequence D | Very high radiation with medium to strong wind Mean temperature: 26°C Maxi temperature: 32 °C |
| Sequence E | Sequence with high global radiation high diffuse radiation Mean temperature: 27°C Maxi temperature: 30°C This sequence follow a very rainy sequences. |
| Sequence F | Sequence of rainy days. This kind of situations is relatively frequent in the wet season Low radiation Mean temperature: 26°C Maxi temperature: 30°C |
| Sequence G | Dry and hot sequence. This kind of sequence is relatively rare in the wet season, however, we used it because it occurs a maximal temperature of 35°C. High radiation Mean temperature: 27°C Medium wind (3 to 6 m.s-1) |

For the sequence G, we found that for dry conditions, the most important sensible cooling capacity (22 KWh) for a maximal temperature of 35°C for the day. We found the same result as for sequence A, but for a lower maximal temperature

(30° C). It must be caused by the difference of wind speed between these two sequences. Effectively, sequence A is less windy as sequence G.

Table 6. Cooling capacities for the different sequences.

| Type of sequence | Nature of the daily capacity | Sensible cooling capacity (Kwh) | Latent cooling capacity (Kwh) | Total cooling capacity (Kwh) |
|---|---|---|---|---|
| Sequence A | MEAN | 18.5 | 2.6 | 21.2 |
| | MAX | 21 | 3.5 | 24.5 |
| Sequence B | MEAN | 10.5 | 1.8 | 12.3 |
| | MAX | 11.5 | 2.8 | 14.3 |
| Sequence C | MEAN | 18.5 | 2.5 | 21 |
| | MAX | 19.8 | 3 | 22.8 |
| Sequence D | MEAN | 13.5 | 2.5 | 16 |
| | MAX | 18 | 4.7 | 22.7 |
| Sequence E | MEAN | 14.3 | 2.6 | 16.9 |
| | MAX | 15 | 6 | 21 |
| Sequence F | MEAN | 10 | 4 | 14 |
| | MAX | 15.5 | 6 | 21.5 |
| Sequence G | MEAN | 18.5 | 3 | 21.5 |
| | MAX | 21 | 3.4 | 24.4 |

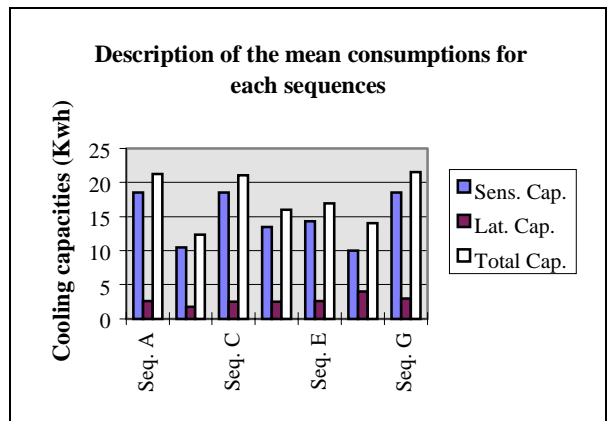

fig. 7: Graphical description of the mean cooling capacities.

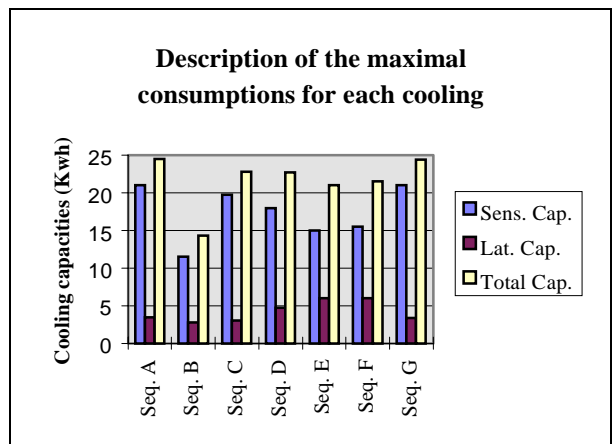

fig. 8: Graphical description of the mean cooling capacities.

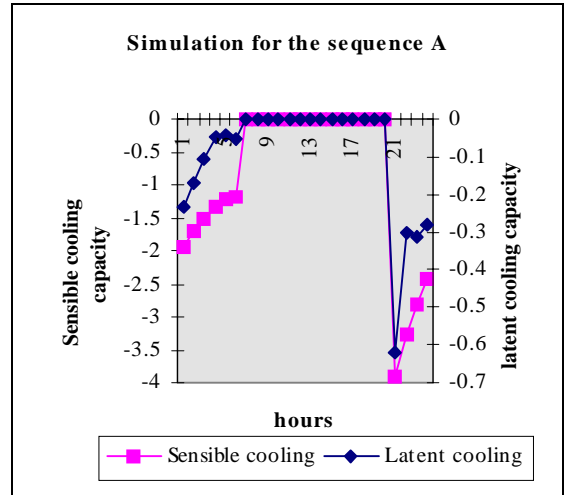

fig. 9: Example of hourly demand of the simulation.

## CONCLUSION.

The aim of this paper was to define the methodology use to determine weather sequences for buildings simulation. Specific study of the cooling capacity were made. The weather sequences were chose according to their «history», and specific criteria based on professional needs. As we have no reference year on the island, we just showed that the choice of the sequence could lead to very various results and. We also showed that the worst conditions appear for mean temperature, high radiation and above all, breeze and high relative humidity. The simulations also showed the great latent consumption for the sequences after rainy days. This kind of simulation is useful to trained the buildings under extreme climatic conditions. The next stage of this study should be to compare our results with this obtained by using a typical meteorological year or with a weather data simulator such as the one used in the software TRNSYS [19]. This method was useful in simulating the operation of the buildings under extreme climatic conditions. It would be interesting then to study the methodology used by Haberl and al. then to compare the results.

## ACKNOWLEDGMENTS

This research is supported by a grant from the Conseil Regional of Reunion Island. The weather data were supplied by the Meteo France services in Reunion Island.